\documentstyle[12pt]{article}

\font\tenmsb=msbm10 scaled\magstep 1
\font\sevenmsb=msbm7 scaled \magstep 1
\font\faivemsb=msbm5 scaled \magstep 1
\newfam\msbfam
\textfont\msbfam=\tenmsb
\scriptfont\msbfam=\sevenmsb
\scriptscriptfont\msbfam=\faivemsb\def\Bbb#1{{\fam\msbfam #1}}

\font\tengothic=eufm10 scaled\magstep 1
\font\sevengothic=eufm7 scaled\magstep 1
\newfam\gothicfam
\textfont\gothicfam=\tengothic
\scriptfont\gothicfam=\sevengothic

\newcommand{\be}{\begin{equation}}
\newcommand{\ee}{\end{equation}}
\newcommand{\dlt}{\delta}
\newcommand{\ra}{\rightarrow}
\newcommand{\vp}{\varphi}
\newcommand{\bt}{\beta}
\newcommand{\al}{\alpha}
\newcommand{\prt}{\partial}

\newcommand{\om}{\omega}
\newcommand{\gm}{\gamma}
\newcommand{\ep}{\varepsilon}
\newcommand{\sgm}{\sigma}

\begin{document}

\begin{center}
{\Large{\bf Summation of Power Series by Self-Similar Factor 
Approximants} \\ [5mm]
V.I. Yukalov$^{1*}$, S. Gluzman$^2$, and D. Sornette$^{2,3,4}$} \\ [3mm]
{\it
$^1$Bogolubov Laboratory of Theoretical Physics, \\
Joint Institute for Nuclear Research, Dubna 141980, Russia\\ [5mm]
$^2$Institute of Geophysics and Planetary Physics, \\
University of California, Los Angeles, California 90095 \\ [5mm]
$^3$ Department of Earth and Space Science,\\
University of California, Los Angeles, California 90095 \\ [5mm]
$^4$ Laboratoire de Physique de la Mati\`ere Condens\'ee, \\
CNRS UMR6622  and Universit\'e des Sciences, \\
Parc Valrose, 06108 Nice Cedex 2, France}

\end{center}

\vskip 2cm

\begin{abstract}

A novel method of summation for power series is developed. The method
is based on the self-similar approximation theory. The trick employed is
in transforming, first, a series expansion into a product expansion and
in applying the self-similar renormalization to the latter rather to the
former. This results in self-similar factor approximants extrapolating
the sought functions from the region of asymptotically small variables
to their whole domains. The method of constructing crossover formulas,
interpolating between small and large values of variables is also analysed.
The techniques are illustrated on different series which are typical of 
problems in statistical mechanics, condensed-matter physics, and, generally, 
in many-body theory.

\end{abstract}

\vskip 1.5cm


\vskip 1.5cm

{\parindent=0pt
{\it Keywords}: Summation of power series; Self-similar approximation theory;
Extrapolation and interpolation methods; Critical point phenomena}

\vskip 1cm

{\parindent=0pt
$^*$corresponding author 

yukalov@thsun1.jinr.ru}

\newpage

\section{Introduction}

It is the standard situation for practically all realistic problems in physics
and applied mathematics that their solutions are obtained by some kind of
perturbation theory or iterative procedure resulting in asymptotic series in
powers of a parameter or a variable. Such series are usually only asymptotic
and diverge for finite values of the expansion variable, while what is of
physical interest are exactly the finite values of the latter. That is why
the problem of defining an effective sum of divergent asymptotic series is
of paramount importance. The most often employed summation techniques are
Pad\'e and Borel summations [1,2]. These techniques have a number of
limitations, because of which in many cases they are either of bad accuracy
or often are not applicable at all.

An alternative approach to the problem of defining effective sums of asymptotic
series or effective limits of iterative sequences has been developed, being
based on the ideas of optimal control theory, renormalization group theory,
and general dynamical theory. The principal concepts of this approach are as
follows.

The first pivotal step is the introduction of {\it control functions} whose
role is to govern an optimal convergence of approximation sequences [3]. This
results in the {\it optimized perturbation theory} which has been widely
employed for a variety of applications (see surveys in [4,5]).

The second idea is to consider the passage from one successive approximation
to another as the motion on the manifold of approximants, where the approximant
order plays the role of discrete time. The recurrent relations, representing
this motion, are formalized by means of {\it group self-similarity} [6--11].

A dynamical system in discrete time, whose trajectory is bijective to an
approximation sequence, is called the {\it approximation cascade}. Embedding
the latter into an approximation flow makes it possible to derive differential
and integral equations of motion, whose fixed points represent {\it 
self-similar approximants} to the sought function [6--11]. The stability 
of the calculational procedure is characterized by {\it local multipliers} 
[8--11].

Another key point of the approach is the introduction of control functions
with the help of {\it fractal transforms} [12--14]. This allows one to analyse
asymptotic series by transforming them into their fractal counterparts and
then invoking all the machinery of the self-similar approximation theory
for the transformed series. In this way, the approximants, possessing a nice
self-similar structure, were obtained, such as the {\it self-similar exponential
approximants} [13--15] and {\it self-similar root approximants} [16--18].

In the present paper, we suggest and give a detailed analysis of one more type
of approximants possessing a nice self-similar structure, which, because of
their form, we name the {\it self-similar factor approximants}. In Section 2,
we explain the origin of these approximants and discuss some variants of their
usage. Then, in Section 3, we illustrate the method on several functions
containing exponentials. Section 4 contains examples of more complicated
and more realistic physical problems. Quartic anharmonic models are treated
in section 5. Calculations of critical points and critical indices for some
critical phenomena are presented in Section 6. The results are summarized in
Section 7, where further possible developments of the method are discussed.

\section{Method of self-similar factor approximants}

Suppose we are looking for a physical quantity described by a real function
$f(x):\; \Bbb{R}\ra\Bbb{R}$. It is the common case for realistic physical
problems that the sought function is defined through so complicated equations
that the problem can be solved only approximately, by invoking a kind of
perturbation theory. Assume that the latter yields perturbative approximants
\be
\label{1}
f(x) \simeq f_k(x) \qquad (x\ra 0) \; ,
\ee
of order $k=0,1,2,\ldots$, in the asymptotic vicinity of $x=0$. For a while,
the concrete meaning of the function $f(x)$ and of the variable $x$ is not
important. In particular applications, $x$ may be a spatial variable, a 
coupling parameter, or any other physical parameter. This may also be a time 
variable in expansions having to do with differential evolution equations 
[19--21].

We may assume, without loss of generality, that the sought function $f(x)$
has the property $f(0)=1$, so that $f_k(0)=1$. As is evident, this can be
achieved for any function by the appropriate scaling and normalization. Then
an asymptotic series in  powers of $x$ takes the form
\be
\label{2}
f_k(x) = 1 + \sum_{n=1}^k a_n\; x^n \; .
\ee
If we apply the technique of the self-similar approximation theory to the
fractal transform of series (2), we would come, depending on the choice of
control functions, to the self-similar exponential or root approximants
[13--18]. But here, we follow another route.

We may notice [22] that each series (2) can always be equivalently presented
as a product
\be
\label{3}
f_k(x) =\prod_{i=1}^k (1 + b_i x) \; ,
\ee
with the coefficients $b_i$ defined by the accuracy-through-order relations
with respect to series (2). Now, we apply the procedure of self-similar
renormalization not to the whole series (2) but separately to each factor
of the product (3). This results in the transformation
\be
\label{4}
1+b_i x \ra (1 + A_i x)^{n_i} \; ,
\ee
in which $A_i$ and $n_i$ are control parameters (see details in [12--18]).
In this way, starting with product (3), we come to the {\it self-similar
factor approximants}
\be
\label{5}
f_k^*(x) \equiv \prod_{i=1}^k ( 1+ A_i x)^{n_i} \; .
\ee
Control parameters $A_i$ and $n_i$ can be found by means of the re-expansion
procedure [23], expanding expression (5) in powers of $x$ and equating the 
like terms from series (2). It is also possible to define some of the 
controllers by matching the behaviour of the factor approximant (5) with 
the large-$x$ asymptotic behaviour of $f(x)$, provided this is available. 
For instance, if the asymptotic form

\be
\label{6}
f(x) \simeq f_\infty\; x^\bt \qquad (x\ra\infty)
\ee
is known, then we may require the validity of the crossover condition
\be
\label{7}
\prod_{i=1}^k A_i^{n_i} = f_\infty \; , \qquad \sum_{i=1}^k n_i =\beta \; .
\ee
The resulting crossover formula (5) would sew together two asymptotic
expressions, at small $x\ra 0$ and large $x\ra\infty$.

As a zero-order approximant, we may set
\be
\label{8}
f_0^*(x) \equiv 1 + a_1 x \; .
\ee
The first-order approximant (5) is
\be
\label{9}
f_1^*(x) = ( 1 + A_1 x)^{n_1} \; ,
\ee
and for the second-order, we have
\be
\label{10}
f_2^*(x) = ( 1 + A_1)^{n_1}( 1+ A_2 x)^{n_2} \; .
\ee
Thus, we construct approximants of any order. For each approximant $f_k^*(x)$,
the accuracy-through-order relation defines its own controllers $A_i=A_{ik}$
and $n_i=n_{ik}$. However, to avoid cumbersome notation, we shall not employ
the double indexation, keeping in mind that the controllers $A_i$ and $n_i$
are a sort of variables that are different for different approximation orders
$k=1,2,\ldots$.

Each self-similar factor approximant (5) of order $k$ contains $2k$ controllers
$A_1,\ldots,A_k$ and $n_1,n_2,\ldots,n_k$. To define these through the
re-expansion procedure, one needs to have $2k$ nontrivial terms of series (2).
Thus, an even-order perturbative series (2) uniquely defines the corresponding
factor approximant (5). The possibility of using odd-order series will be
discussed in Section 7.

It may happen that, calculating the controllers $A_i$ and $n_i$, we shall
obtain for some of them complex values. If the sought function is real, then
the complex-valued controllers should come in complex conjugate pairs, so
that approximant (5) be real. This means that the latter may contain the
factors of the type $|z^\al|^2$, with both $z$ and $\al$ being complex valued.
Since
$$
|z^\al| = | z|^{{\rm Re}\;\al}\;
\exp\left \{ -({\rm Im}\;\al)\; {\rm arg}\; z\right \}
$$
is a nonalgebraic function of $z$, hence expression (5) may, in general, be
a nonalgebraic function. According to definition [24], an analytical function
that is not algebraic is termed transcendental.

In this way, formula (5) can represent the following functions. When all
powers $n_i$ are real integers, positive or negative, then $f_k^*(x)$ is
a rational function. In the case when all powers are equal to $\pm 1$,
expression (5) coincides with the form of Pad\'e approximants, which are, 
therefore, just a narrow particular case of the self-similar factor 
approximants. When some of $n_i$ are real but not integer, then 
$f_k^*(x)$ is an irrational function. And if some of $A_i$ and $n_i$ are
complex valued, then $f_k^*(x)$ represents a transcendental function. As far
as any sought function, having the same form (5), is exactly reproducible by
$f_k^*(x)$, whose controllers are defined by the re-expansion procedure [23],
this means that the class of functions, that can be exactly reconstructed by
the self-similar factor approximants, is rather wide, including rational,
irrational, and some transcendental functions. This class is essentially
wider than that related to Pad\'e approximants and consisting solely of
rational functions. Consequently, the factor approximants (5) can provide
much better accuracy than Pad\'e approximants for the majority of functions
and also can be employed when the latter are not applicable at all.

Approximants (5) can well approximate such entire transcendental functions
as exponentials. This can be easily understood noticing that if a factor
$(1+Ax)^n$ is such that $A=a/n$, where $a$ is a constant and $n\ra\infty$,
then such a factor gives
$$
\left ( 1 +\frac{a}{n}\; x\right )^n \ra e^{ax} \qquad (n\ra\infty) \; ,
$$
that is, it is equivalent to an exponential. Therefore, if in some particular
calculations we find a very small $A$ but a very large $n$, this hints that
the corresponding factor $(1+Ax)^n$ is close to the exponential $\exp(nAx)$.
This is why the factor approximants can provide a reasonable approximation
for exponential functions.

\section{Reconstruction of functions including exponentials}

Exponential functions are ubiquitous is natural sciences representing various
physical laws. At the same time, this is a simple example of an entire
transcendental function. The best way of approximating these functions would
be by means of the self-similar exponentials [13--15], which often reconstruct
exponential-type functions exactly [25]. But the factor approximants, as is
explained in the previous section, should also mimic well exponential behaviour.
This is illustrated in the following examples.

\subsection{Nonmonotonic exponential function with maximum}

We choose a nonmonotonic function in order to stress that such a nonmonotonic
behaviour presents no problem for the method. Let us consider the function
\be
\label{11}
f(x) = \exp\left \{ \frac{x}{(1+x)^{5/2}} \right \} \; .
\ee
Its asymptotic expansion at $x\ra 0$ yields the finite sums (2) with the
coefficients
$$
a_1 = 1 \; , \qquad a_2 = -2 \; , \qquad a_3 = 2.042 \; ,
\qquad a_4=-0.271\; ,
$$
$$
a_5 = -3.572 \; , \qquad a_6 = 8.636 \; , \qquad a_7 = -12.661 \; ,
\qquad a_8=12.183\; ,
$$
and so on. To approximate function (11), we construct the self-similar factor
approximants (5). The first-order approximant (9) is yet too simple to catch 
well the peculiarities of the function variation. So, here and in what 
follows, we shall start the analysis with the second-order approximant (10).

Here for $f_2^*(x)$, we have
$$
A_1=1.658+0.840\; i \; , \qquad A_2 = A_1^* \; ;
$$
$$
n_1=-0.244-1.077\; i \; , \qquad n_2=n_1^* \; .
$$
But, as $x\ra\infty$, the function $f_2^*(x)$ decreases to zero, since 
$n_1+n_2=-0.488$, while function (11) tends to unity.

For $f_3^*(x)$, we get
$$
A_1=1.3540\; , \qquad A_2 =1.0015+0.4034\; i\; , \qquad   A_3^*=A_2^* \; ;
$$
$$
n_1=11.0051\; , \qquad n_2=-5.4336+3.7402\; i\; , \qquad n_3=n_2^* \; .
$$
Note that the enumeration of the controllers $A_i$ and $n_i$ can be arbitrary,
as far as the order of factors in Eq. (5) is interchangeable. As a rule, we 
shall enumerate them according to the descending order of the values 
$|{\rm Re}\; A_i|$. Now, $n_1+n_2+n_3=0.138$, and the approximation is not 
bad till $x\sim 150$.

For $f_4^*(x)$, we find
$$
A_1=1.1455 \; , \qquad A_2 = 1.0076+0.1703\; i \; , \qquad A_3= A_2^* \; ,
\qquad A_4=0.5156\; ;
$$
$$
n_1=84.0147\; , \qquad n_2 =-40.1038+36.6118\; i \; , \qquad n_3=n_2^* \; ,
\qquad n_4=-3.7834 \; .
$$
The sum of the powers decreases, $n_1+n_2+n_3+n_4=0.024$. And the accuracy 
is very good up to $x$ of order of hundred, with the error being less than 
$5\%$.

\subsection{Nonmonotonic exponential function with minimum}

Let us now turn to the function
\be
\label{12}
f(x) = \exp\left \{ -\; \frac{x}{(1+x)^{5/2}} \right \}
\ee
having a minimum. The corresponding coefficients in series (2) are
$$
a_1 = -1 \; , \qquad a_2 = 3 \; , \qquad a_3 = -7.042 \; ,
\qquad a_4=15.354\; ,
$$
$$
a_5 = -32.261 \; , \qquad a_6 = 65.951 \; , \qquad a_7 = -131.714 \; ,
\qquad a_8=257.734\; .
$$
This function also tends to $1$ at $x\ra\infty$.

For $f_2^*(x)$, we have
$$
A_1=1.65790 + 0.83984\; i \; , \qquad A_2 = A_1^* \; ;
$$
$$
n_1=0.24381+1.07665\; i \; , \qquad n_2=n_1^* \; .
$$
This approximant works well till $x\sim 10$, but at $x\ra\infty$ its behaviour
is not correct since $n_1+n_2=0.488$.

The controllers of $f_3^*(x)$ are
$$
A_1=1.35398\; , \qquad A_2 =1.00149+0.40337\; i\; , \qquad   A_3=A_2^* \; ;
$$
$$
n_1=-11.00514\; , \qquad n_2=5.43355 - 3.74022\; i\; , \qquad n_3=n_2^* \; .
$$
This function decreases at $x\ra\infty$, since $n_1+n_2+n_3=-0.133$. 
Reasonable accuracy is provided till $x\sim 10$.

The next approximant $f_4^*(x)$ has the parameters
$$
A_1=1.14550 \; , \quad A_2 = 1.00759+0.17032\; i \; , \quad A_3= A_2^* \; ,
\quad A_4=0.51563\; ;
$$
$$
n_1=-84.01466\; , \quad n_2 =40.10379-36.61184\; i \; , \quad n_3=n_2^* \; ,
\quad n_4=3.78340 \; .
$$
The function decreases much slower than the previous approximations, since
$n_1+n_2+n_3+n_4=-0.024$. The accuracy is good up to $x\sim 100$. For 
instance, the percentage error at $x=40$ equals $-2\%$.

\subsection{Spectral density of black-body radiation}

This is given by the Planck formula $\rho(x)=Cx^3/(e^x-1)$. This is also 
a nonmonotonic function with a maximum. To get the presentation (2), we 
define $f(x)\equiv\rho(x)/Cx^2$, which results in
\be
\label{13}
f(x) = \frac{x}{e^x-1} \; ,
\ee
where $x\equiv\hbar\om/k_BT$. The coefficients of series (2) are
$$
a_1 = -\; \frac{1}{2}\; , \qquad a_2 =\frac{1}{12} \; , \qquad a_3=0\; , 
\qquad a_4=-\; \frac{1}{720} \; ,
$$
$$
a_5=0\; , \qquad a_6=\frac{1}{30240} \; , \qquad a_7=0 \; , \qquad
a_8=-\; \frac{1}{1209600} \; ,
$$
$$
a_9=0\; , \qquad a_{10}=\frac{1}{47900160}\; , \qquad a_{11}=0 \; ,
\qquad a_{12}=-\; \frac{691}{1307674368000}\; .
$$

For $f_2^*(x)$, we get
$$
A_1=-0.05+0.119024\; i \; , \qquad A_2 = A_1^* \; ;
$$
$$
n_1=-1+2.520504\; i \; , \qquad n_2=n_1^* \; .
$$
The function decreases at $x\ra\infty$, since $n_1+n_2=-2$, but not so fast as
Eq. (13).

For $f_3^*(x)$, one has
$$
A_1=-0.052201\; , \qquad A_2 =-9.613987\times 10^{-3}+0.150710\; i\; , \qquad
A_3 = A_2^* \; ;
$$
$$
n_1=7.654768\; , \qquad n_2=-1.327382+0.417823\; i\; , \qquad n_3=n_2^* \; .
$$
The function increases at $x\ra\infty$ because $n_1+n_2+n_3=5$, which is not
correct.

The controllers of $f_4^*(x)$ are
$$
A_1=-0.026326 +0.044214\; i \; , \quad A_2=A_1^* \; , \quad
A_3= -1.452250\times 10^{-3}+0.158034\; i \; ,\quad A_4=A_3^*\; ;
$$
$$
n_1=-0.939416+5.998214\;i\; , \quad n_2 =n_1^* \; , \quad
n_3=-1.060584+0.070005\; i \; , \quad n_4=n_3^* \; .
$$
The sum $n_1+n_2+n_3+n_4=-4$ shows that the function decreases as $x\ra\infty$,
in agreement with the behaviour of Eq. (13).

To analyse the behaviour of higher-order approximants, we have also 
calculated $f_5^*(x)$, for which $\sum_{k=1}^5 n_k=0.367295$; because of 
this $f_5^*(x)$ increases at infinity, though not as fast as $f_3^*(x)$. And
for $f_6^*(x)$, we find $\sum_{k=1}^6 n_k=-3.300722$; hence this approximant 
decreases as $x\ra\infty$. With increasing order $k$, the interval of $x$, 
where $f_k^*(x)$ provides a very good approximation to $f(x)$, becomes 
larger. Figure 1 presents the difference $f_4^*(x)-f(x)$.

\subsection{Specific heat of diatomic gas}

The oscillational part of the specific heat for a diatomic gas has the
form [26]
\be
\label{14}
C(x) = \frac{x^2\; e^x}{(e^x-1)^2} \; ,
\ee
with $x\equiv\hbar\om/k_BT$. The expansion coefficients of series (2) are
$$
a_1 = -\; \frac{1}{12}\; , \qquad a_2 =\frac{1}{240} \; , \qquad
a_3=-\; \frac{1}{6048}\; , \qquad a_4=-\; \frac{1}{172800} \; ,
$$
$$
a_5=-\; \frac{1}{5322240} \; , \qquad a_6=\frac{691}{118879488000} \; ,
$$
and so on. Expression (14) tends to zero as $x\ra\infty$.

For the second approximant $C_2^*(x)$, we find
$$
A_1=0.0251 \; , \qquad A_2 = 0.0026 \; ;
$$
$$
n_1=-2.0656 \; , \qquad n_2=-11.9344 \; .
$$
The approach to zero, as $x\ra\infty$, is sufficiently fast, since 
$n_1+n_2=-14$.

For $C_3^*(x)$, one gets
$$
A_1=0.025148\; , \qquad A_2 =198.278139\; , \qquad 
A_3=2.629939\times 10^{-3} \; ;
$$
$$
n_1=-2.065650\; , \qquad n_2= 0 \; , \qquad n_3=-11.934396 \; .
$$
Here, $n_1+n_2+n_3=-14$. Both $C_k^*(x)$ are of good accuracy, with the 
difference $C_k^*(x)-C(x)$ less than $10^{-4}$. Note that small $A_3$ with 
large $n_3$ hint at the exponential dependence. The comparison of $C_3^*(x)$ 
with $C(x)$, given in Eq. (14), shows that this factor approximant is 
essentially more accurate than the best Pad\'e approximants $P_{[4/2]}(x)$
and $P_{[5/1]}(x)$, as is illustrated in Fig. 2.

\subsection{Regular solution from Debye-Huckel theory}

Consider the function
\be
\label{15}
f(x) = \frac{2}{x} \; - \; \frac{2}{x^2}\; \left ( 1 - e^{-x} \right )
\ee
appearing in the Debye-Huckel theory [26]. This function contains an 
exponential, but the large-$x$ asymptotic behaviour is of power law,
$$
f(x) \simeq 2 \left ( \frac{1}{x}\; - \; \frac{1}{x^2}\right )
\qquad (x\ra \infty) \; .
$$
The coefficients of series (2) are
$$
a_1 = -\; \frac{1}{3}\; , \qquad a_2 =\frac{1}{12} \; , \qquad
a_3=-\; \frac{1}{60} , \qquad a_4= \frac{1}{360} \; ,
$$
$$
a_5=-\; \frac{1}{2520} , \qquad a_6=\frac{1}{20160} \; , \qquad
a_7=-\; \frac{1}{181440} , \qquad a_8= \frac{1}{1814400} \; .
$$
Constructing the factor approximants (5), we shall invoke the crossover 
condition (7).

For $f_2^*(x)$, we have
$$
A_1=0.177337+0.176694\; i \; , \qquad A_2 = A_1^* \; ;
$$
$$
n_1=-0.5+0.441432\; i \; , \qquad n_2=n_1^* \; .
$$
The maximal error of this approximant is about $2.5\%$ at $x\approx 15$.

The controllers of $f_3^*(x)$ are
$$
A_1=0.153026\; , \qquad A_2 =0.076787+0.156796\; i\; , \qquad
A_3=A_2^* \; ;
$$
$$
n_1=-1.146332\; , \qquad n_2=0.073166+0.539398\; i\; , \qquad n_3=n_2^* \; .
$$
The maximal error now is around $-1\%$ at $x\approx 20$. Percentage errors 
of $f_3^*(x)$ and the two-point Pad\'e approximant $P_{[3/2]}(x)$ are shown 
in Fig. 3.

For $f_4^*(x)$, we get
$$
A_1=0.095338 +0.056647\; i \; , \quad A_2=A_1^* \; , \quad
A_3= 0.043027+0.139489\; i \; ,\quad A_4=A_3^*\; ;
$$
$$
n_1=-0.966783+0.967772\;i\; , \quad n_2 =n_1^* \; , \quad
n_3=0.466783+0.285025 \; i \; , \quad n_4=n_3^* \; .
$$
Here the maximal, with respect to all $x\in[0,\infty)$, error is about $0.5\%$
at $x\approx 30$. The best two-point Pad\'e approximant [1] involving the same
number of terms has the maximal error $-6\%$ at $x\approx 20$, which is an 
order larger than the maximal error of $f_4^*(x)$. The accuracy of the latter 
may be improved even more, if to take into account two terms from the 
large-$x$ expansion. Then the maximal error can be reduced to $0.15\%$.

\section{Structure factors and radiation intensity}

The quantities such as structure factors and radiation intensity are directly
observable in experiments. That is why much attention is paid to their
theoretical description. Below we show how one can derive rather simple and
quite accurate expressions for such quantities, by means of the self-similar
factor approximants.

\subsection{Structure factor of branched polymers}

The structure factor $S(x)$ is usually expressed in scaled units as a function
of $x\equiv k^2R^2/d$, where $k$ is a wave vector, $R$ is gyration radius, and
$d$ is spatial dimensionality [27,28]. For $d=3$, the structure factor of
branched polymers is given [28] by a confluent hypergeometric function 
$S(x)=F_1(1;3/2;3x/2)$. The behaviour at large $x$ is
$$
S(x) \simeq \frac{1}{3}\; x^{-1} \qquad (x\ra\infty) \; .
$$
At small $x$, the coefficients of series (2) are
$$
a_1=-1\; , \qquad a_2=0.6\; , \qquad a_3=-0.257 \; .
$$
We shall use the crossover condition (7). Then for $S_2^*(x)$, we find
$$
A_1=0.247387+0.483295\; i \; , \qquad A_2=A_1^*\; ; 
$$
$$
n_1=-0.5-0.778627\; i \; , \qquad n_2=n_1^* \; .
$$
The approximant $S_2^*(x)$ agrees well with the exact $S(x)$, as is shown in
Fig. 4.

\subsection{Structure factor of ring polymers}

The structure factor of three-dimensional dilute ring polymers $S(x)$ can be
considered as a function of $x\equiv (kR)^2$, where $k$ is wave vector and $R$
is gyration radius [29]. In series (2), we have
$$
a_1 = -\; \frac{1}{3}\; , \qquad a_2 =0.0610\; , \qquad a_3 = -0.0073 \; .
$$
Knowing the large-$x$ asymptote
$$
S(x) \simeq 0.63\; x^{-1/2\nu} \qquad (\nu=0.588) \; ,
$$
we shall employ the crossover condition (7). Then we get for $S_2^*(x)$
$$
A_1=0.065483+0.185055\; i \; , \qquad A_2=A_1^*\; ; 
$$
$$
n_1=-0.425170+0.750184\; i \; , \qquad n_2=n_1^*\; .
$$
This gives a rather good accuracy, as compared to numerical results [29].

\subsection{Luminescent intensity of donor-acceptor recombination}

The expression for the temporal dependence of the luminescence intensity, 
arising from radiative recombination of donor-acceptor pairs, has the 
form [30,31] 
$$
I(z) = - \exp\{ aK(z)\} \; \frac{d}{dz}\; K(z) \; ,
$$
in which $z\equiv wt$, $w$ is a transition probability per unit time, $t$ is 
time, $a$ is a constant, and
$$
K(z) \equiv
\int_0^\infty x^2\left [ \exp\left ( -ze^{-x}\right ) -1 \right ] \; dx \; .
$$
We shall consider the reduced quantity
$$
f(z) \equiv \frac{I(z)}{I(0)} \; , \qquad I(0) = 2 \; ,
$$
with the typical parameter $a=6.514\times 10^{-7}$. Then the coefficients of 
short-time series (2) are
$$
a_1=-0.125\; , \qquad a_2=0.019\; , \qquad a_3=-2.604\times 10^{-3}\; ,
$$
$$
a_4=3.334\times 10^{-4}\; , \qquad a_5=-3.858\times 10^{-5}\; , \qquad
a_6=4.050\times 10^{-6}\; .
$$
In the long-time limit, one has
$$
f(z) \simeq \frac{(\ln z)^2}{4z}\; \exp\left \{ -\; \frac{a}{3}\;
(\ln z)^3\right \} \qquad (z\ra\infty) \; ,
$$
which tends to zero.

For $f_2^*(z)$ we find
$$
A_1=0.132653+0.072661\; i \; , \quad A_2=A_1^*\; ; 
$$
$$
n_1=-0.256845+0.391265\; i \; , \quad n_2=n_1^*\; .
$$
This approximant tends to zero, as $z\ra\infty$, since $n_1+n_2=-0.514$.

For $f_3^*(z)$, we obtain
$$
A_1=0.098681\; , \qquad A_2 =0.087209+0.090440\; i\; , \qquad
A_3=A_2^* \; ;
$$
$$
n_1=-1.292027\; , \qquad n_2=0.293510+0.269217\; i\; , \qquad n_3=n_2^* \; .
$$
This also tends to zero, as $z\ra\infty$, with the power law $n_1+n_2+n_3
=-0.705$. In the range $0\leq z\leq 10^3$, the maximal error is about $2\%$ 
at $z\approx 200$. The best Pad\'e approximant $P_{[3/3]}(z)$ is much less 
accurate, as is seen in Fig. 5.

\section{Models of quartic anharmonic oscillators}

Anharmonic oscillator models, especially low-dimensional ones, are usually
considered as touchstones for checking the validity of any approximation
technique. This is because these models mimic the main mathematical features
of more realistic, but much more complicated, problems in many-body theory 
and field theory. Below, we illustrate the applicability of the method of 
self-similar factor approximants to two low-dimensional anharmonic models.

\subsection{Partition function of zero-dimensional model}

The partition function is modelled by the integral
\be
\label{16}
Z(g) = \frac{1}{\sqrt{\pi}} \; \int_{-\infty}^{+\infty}
\exp\left ( -\vp^2-g\vp^4 \right ) \; d\vp \; ,
\ee
in which $g$ imitates a coupling parameter. Coefficients of series (2) are
given by the formula
$$
a_n =\frac{(-1)^n}{\sqrt{\pi}\; n!} \; \Gamma\left (2n+
\frac{1}{2}\right ) \; .
$$
In the strong-coupling limit, one has
\be
\label{17}
Z(g) \simeq 1.022765\; g^{-1/4} \qquad (g\ra\infty) \; .
\ee

Following the technique of Section 2, we find for $Z_2^*(g)$ the controllers
$$
A_1=19.141 \; , \qquad A_2= 4.859 \; ; 
$$
$$
n_1=-0.00862 \; , \qquad n_2=-0.120\; .
$$
At large $g$, this yields
$$
Z_2^*(g) \simeq 0.806\; g^{-0.129} \qquad (g\ra\infty) \; .
$$
The approximant gives a reasonable accuracy at finite $g$ and is qualitatively
correct at large $g$, though the decrease of the partition function (16) is not
as fast as in Eq. (17). The amplitude of the decay is predicted with an error
$-21\%$ and the power, with an error $48\%$.

For $Z_3^*(g)$, we get
$$
A_1=31.220\; , \qquad A_2 =13.317\; , \qquad A_3=3.464 \; ;
$$
$$
n_1=-0.000526\; , \qquad n_2=-0.022\; , \qquad n_3=-0.125 \; .
$$
In the strong-coupling limit,
$$
Z_4^*(g) \simeq 0.807\; g^{-0.148} \qquad (g\ra \infty) \; .
$$
Here the prediction errors are $-21\%$ for the amplitude and $41\%$ for the
power.

The approximant $Z_4^*(g)$ contains
$$
A_1=43.965 \; , \qquad A_2=23.064 \; , \qquad
A_3= 10.294\; ,\qquad A_4=2.677\; ;
$$
$$
n_1=-0.000025\; , \qquad n_2 =-0.00259\; , \qquad
n_3=-0.035 \; , \qquad n_4=-0.124 \; .
$$
The strong-coupling limit gives
$$
Z_4^*(g)\simeq 0.810\; g^{-0.161} \qquad (g\ra \infty) \; .
$$
The errors are $-21\%$ for the amplitude and $36\%$ for the power index.

For $Z_5^*(g)$, we find
$$
A_1=57.165211 \; , \qquad A_2=33.720143 \; , \qquad
A_3= 18.543443\; ,
$$
$$
A_4=8.396565\; , \qquad A_5=2.174639\; ;
$$
$$
n_1=-1.025651 \times 10^{-6}\; , \qquad n_2 =-2.150967\times 10^{-4} \; ,
\qquad n_3=-6.028095\times 10^{-3} \; ,
$$
$$
n_4=-0.044164 \; , \qquad n_5=-0.119596\; .
$$
The strong-coupling limit is
$$
Z_5^*(g) \simeq 0.81445\; g^{-0.170} \qquad (g\ra \infty) \; .
$$
Here, the errors of the amplitude is $-20\%$ and that of the power index
is $32\%$.

And for $Z_6^*(g)$, we obtain
$$
A_1=66.087061 \; , \qquad A_2=41.445220 \; , \qquad
A_3= 25. 096672\; ,
$$
$$
A_4=13.752049\; , \qquad A_5=6.067701\; , \qquad A_6=1.511098 ;
$$
$$
n_1=-1.153412 \times 10^{-7}\; , \qquad n_2 =-3.352991\times 10^{-5} \; ,
\qquad n_3=-1.330855\times 10^{-3} \; ,
$$
$$
n_4=-0.014434 \; , \qquad n_5=-0.057625\; , \qquad n_6=-0.110557 \; .
$$
In the strong-coupling limit, this leads to
$$
Z_6^*(g) \simeq 0.82548 \; g^{-0.184} \qquad (g\ra \infty) \; .
$$
The errors are $-19\%$ for the amplitude and $26\%$ for the index. These 
results demonstrate monotonic numerical convergence.

The accuracy of approximants can be essentially improved by invoking the 
crossover condition (7). Then for $Z_2^*(g)$, we have
$$
A_1=8.57593 \; , \qquad A_2= 0.30987 \; ; \qquad
n_1=-0.08136 \; , \qquad n_2=-0.16864\; ;
$$
$$
A_1^{n_1}\; A_2^{n_2} =1.023 \; , \qquad n_1+n_2=-0.25 \; .
$$
This approximant provides a good accuracy in the whole region of 
$g\in[0,\infty)$, with the maximal error of $6\%$ at $g\approx 5$.

For $Z_3^*(g)$, we find
$$
A_1=19.47265\; , \qquad A_2 =5.21916\; , \qquad A_3=0.18018 \; ;
$$
$$
n_1=-7.94804\times 10^{-3}\; , \qquad n_2=-0.10947\; , \qquad 
n_3=-0.13258 \; ;
$$
$$
A_1^{n_1}\; A_2^{n_2}\; A_3^{n_3} =1.023\; , \qquad n_1+n_2+n_3=-0.25 \; .
$$
The maximal error is about $4\%$ at $g\approx 10$.

The next approximant $Z_4^*(g)$ contains
$$
A_1=31.45590 \; , \qquad A_2=13.55695 \; , \qquad
A_3= 3.72340\; ,\qquad A_4=0.12535\; ;
$$
$$
n_1=-4.96209\times 10^{-4} \; , \qquad n_2 =-0.02119\; , \qquad
n_3=-0.11631 \; , \qquad n_4=-0.11200 \; ;
$$
$$
A_1^{n_1}\; A_2^{n_2}\; A_3^{n_3}\; A_4^{n_4} =1.023\; , \qquad 
n_1+n_2+n_3+n_4=-0.25 \; .
$$
The maximal error is around $3.5\%$ at $g\approx 10$.

And for $Z_5^*(g)$, we obtain
$$
A_1=44.148724 \; , \qquad A_2=23.247563 \; , \qquad A_3= 10.482932\; ,
$$
$$
A_4=2.878801\; , \qquad A_5=0.095481 ;
$$
$$
n_1=-2.395229 \times 10^{-5}\; , \qquad n_2 =-2.477078\times 10^{-3} \; ,
\qquad n_3=-0.033250 \; ,
$$
$$
n_4=-0.115813 \; , \qquad n_5=-0.098436\; .
$$
Now the maximal error is close to $3\%$ at $g\approx 20$.

Finally, for $Z_6^*(g)$, we have
$$
A_1=57.315168 \; , \qquad A_2=33.870044 \; , \qquad A_3= 18.694023\; ,
$$
$$
A_4=8.552264\; , \qquad A_5=2.338677\; , \qquad A_6=0.076825\; ;
$$
$$
n_1=-9.879942 \times 10^{-7}\; , \qquad n_2 =-2.072431\times 10^{-4} \; ,
\qquad n_3=-5.811092\times 10^{-3} \; ,
$$
$$
n_4=-0.042533 \; , \qquad n_5=-0.112765\; , \qquad n_6=-0.088682 \; .
$$
For this approximant, the maximal error is about $2.5\%$ at $g\approx 30$.

\subsection{Ground-state energy of {\bf the} anharmonic oscillator}

We consider the one-dimensional anharmonic quartic oscillator with the 
Hamiltonian
\be
\label{18}
H =-\; \frac{1}{2}\; \frac{d^2}{dx^2} + \frac{1}{2}\; x^2 + gx^4 \; ,
\ee
where $x\in(-\infty,+\infty)$. Perturbation theory gives [32] the ground-state
energy $E(g)$ as the series
\be
\label{19}
E(g) = \sum_{n=0}^\infty a_n\; g^n \; ,
\ee
where the first eleven coefficients are
$$
a_0 = 0.5\; , \qquad a_1 = 0.75\; , \qquad a_2 =-2.625\; , 
\qquad a_3 =20.8125\; , \qquad a_4 =-241.2890625\; ,
$$
$$
a_5 = 3580.98046875\; , \qquad a_6 =-63982.8134766\; , 
\qquad a_7 =1329733.72705\; ,
$$
$$
a_8=-31448214.6928\; , \qquad a_9=833541603.263\; , \qquad 
a_{10}=-24478940702.8\; .
$$
Since here $a_0=1/2$, the reduced function $f(g)$, for which $f(0)=1$, has 
to be defined as
$$
f(g) \equiv \frac{E(g)}{E(0)} = 2E(g) \; ,
$$
with the corresponding change of the expansion coefficients in series (2). The
strong-coupling limit of the ground state energy is

\be
\label{20}
E(g) \simeq 0.667986\; g^{1/3} \qquad (g\ra \infty) \; .
\ee

For the self-similar factor approximant $E_2^*(g)$, we have
$$
A_1=17.5973 \; , \qquad A_2= 5.3122 \; ; 
$$
$$
n_1=0.0221 \; , \qquad n_2=0.2091\; .
$$
The large-$g$ limit is
$$
E_2^*(g) \simeq 0.7554\; g^{0.2312} \qquad (g\ra\infty) \; ,
$$
Comparing the latter with the asymptotic form (20), we see that the error of
the amplitude here is $13\%$ and that of predicting the power is $-31\%$.

In the case of $E_3^*(g)$, we get
$$
A_1=26.7402\; , \qquad A_2 =12.4688\; , \qquad A_3=3.8380 \; ;
$$
$$
n_1=1.8017\times 10^{-3}\; , \qquad n_2=0.0547\; , \qquad n_3=0.2005 \; .
$$
The strong-coupling limit gives
$$
E_3^*(g) \simeq 0.7562\; g^{0.2570} \qquad (g\ra\infty) \; .
$$
Here the error of the amplitude is again $13\%$, while that of the power is 
$-23\%$. If we compare $E_3^*(g)$ with the Pad\'e approximant $P_{[3/3]}(g)$, 
providing the best accuracy at finite $g$, then we see that the latter is 
wrong at large $g$, tending to a constant, as $g\ra\infty$, instead of 
increasing.

For $E_4^*(g)$, we find
$$
A_1=36.27784 \; , \qquad A_2=20.11620 \; , \qquad
A_3= 9.69400\; ,\qquad A_4=2.98137\; ;
$$
$$
n_1=1.05387\times 10^{-4} \; , \qquad n_2 =8.41555\times 10^{-3}\; , \qquad
n_3=0.08068 \; , \qquad n_4=0.18273 \; .
$$
In the strong-coupling limit, we have
$$
E_4^*(g) \simeq 0.75229\; g^{0.27193} \qquad (g\ra\infty) \; .
$$
The errors are around $13\%$ for the amplitude and $-18\%$ for the power.

The approximant $E_5^*(g)$ contains
$$
A_1=46.160836 \; , \qquad A_2=28.182719 \; , \qquad A_3= 16.332980\; ,
$$
$$
A_4=7.924575\; , \qquad A_5=2.422310 ;
$$
$$
n_1=4.992768 \times 10^{-6}\; , \qquad n_2 =8.787692\times 10^{-4} \; ,
\qquad n_3=0.018549 \; ,
$$
$$
n_4=0.097564 \; , \qquad n_5=0.164673\; .
$$
The strong-coupling limit yields
$$
E_5^*(g) \simeq 0.747711\; g^{0.28167} \qquad (g\ra\infty) \; .
$$
Now the error of the amplitude is $12\%$ and that of the power is $-15\%$.
There is a monotonic numerical convergence to exact results.

The accuracy of the approximants $E_k^*(g)$ can be essentially improved
with the help of the crossover condition (7). Invoking the latter, we get
for $E_2^*(g)$
$$
A_1=9.457715\; , \qquad A_2=0.869296\; ; 
$$
$$
n_1=0.140915\; , \qquad n_2=-0.192419\; ;
$$
$$
\frac{1}{2}\; A_1^{n_1}\; A_2^{n_2} =0.667986\; , \qquad
n_1+n_2=\frac{1}{3} \; .
$$

For $E_3^*(g)$, we find
$$
A_1=18.112324\; , \qquad A_2=5.896948\; ; \qquad A_3=0.483681\; ;
$$
$$
n_1=0.019178\; , \qquad n_2=0.184859\; , \qquad n_3=0.129296 \; ;
$$
$$
\frac{1}{2}\; A_1^{n_1}\; A_2^{n_2}\; A_3^{n_3}  =0.667986\; ,
\qquad n_1+n_2 +n_3 =\frac{1}{3} \; .
$$

In the approximant $E_4^*(g)$, the controllers are
$$
A_1=27.098877\; , \qquad A_2=12.848678\; ; \qquad A_3=4.253915\; ,
\qquad A_4=0.322173 \; ;
$$
$$
n_1=1.616412 \times 10^{-3}\; , \qquad n_2=0.049729\; , \qquad n_3=0.184752 \; ,
\qquad n_4=0.097236\; ,
$$
with condition (7) being valid.

And for $E_5^*(g)$, we obtain
$$
A_1=36.555891 \; , \qquad A_2=20.393831 \; , \qquad A_3= 9.992627\; ,
$$
$$
A_4=3.302859\; , \qquad A_5=0.236809\; ;
$$
$$
n_1=9.656933 \times 10^{-5}\; , \qquad n_2 =7.776204\times 10^{-3} \; ,
\qquad n_3=0.075166 \; ,
$$
$$
n_4=0.172047 \; , \qquad n_5=0.078248\; ,
$$
again with condition (7) exactly satisfied.

\section{Critical points and critical indices}

The self-similar factor approximants (5) look very appropriate for describing
critical phenomena, where various thermodynamic functions usually possess
power-law behaviour. The direct application of Pad\'e approximants [1] or
their variants [33--35] does not allow for a correct description of critical
indices. But the factor forms, analogous to approximants (5), should provide
an accurate account of typical thermodynamic characteristics in the vicinity
of critical points [36]. Note that critical phenomena exist not only in the
models of statistical mechanics and condensed matter physics but also in the
models of rupture and materials failure [37--40], earthquakes [41,42], and
stock market crashes [43--45]; all such critical phenomena having many common
features [46,47]. Below we show that the self-similar factor approximants
present, probably, the best possibility for characterizing different critical
phenomena.

\subsection{Logarithmic divergence in elliptic integral}

This example illustrates that even when the divergence in a critical point 
is not of power law but logarithmic, it can, anyway, be reasonably well 
described by approximants (5). For this purpose, we consider the elliptic 
integral 
\be
\label{21}
f(t) \equiv \frac{2}{\pi} \int_0^{\pi/2} \frac{dx}{\sqrt{1-t\sin^2x}} \; ,
\ee
which diverges logarithmically at $t_c=1$. The logarithmic divergence implies
that if one defines the critical index as
\be
\label{22}
\al \equiv \lim_{t\ra t_c-0}\; (t-t_c) \frac{d}{dt}\; \ln f(t) \; ,
\ee
then it has to be close to zero. Expanding integral (21) in powers of $t$, one
gets
the asymptotic series of type (2), with the coefficients
$$
a_1=\frac{1}{4}\; , \qquad a_2=\frac{9}{64}\; , \qquad a_3=\frac{25}{256} \; ,
\qquad a_4=\frac{1225}{16384}\; ,
$$
$$
a_5=\frac{3969}{65536}\; , \qquad a_6=\frac{53361}{1048576} \; , \qquad
a_7=\frac{184041}{4194304} \; ,
$$

\vskip 2mm

$$
a_8=\frac{41409225}{1073741824}\; , \qquad a_9=\frac{147744025}{4294967296}\; ,
\qquad a_{10}=\frac{21334223721}{68719476736} \; .
$$

For the approximant $f_2^*(t)$, we have
$$
A_1=-0.9640 \; , \qquad A_2= -0.3485 \; ;
$$
$$
n_1=-0.2218 \; , \qquad n_2=-0.1038\; .
$$
Divergence happens at a critical point $t_c$, with a critical index, defined
in Eq. (22), which are
$$
t_c = \frac{1}{|A_1|} =1.0373\; , \qquad \al=|n_1|=0.222 \; .
$$
The critical point is defined with an error $3.7\%$.

For $f_3^*(t)$, we find
$$
A_1=-0.9839\; , \qquad A_2=-0.6311\; ; \qquad A_3=-0.1767\; ;
$$
$$
n_1=-0.1938\; , \qquad n_2=-0.0721\; , \qquad n_3=-0.0783 \; ;
$$
Divergence is characterized by
$$
t_c=\frac{1}{|A_1|} = 1.0163 \; , \qquad \al=|n_1|=0.194 \; .
$$
Here, the error of $t_c$ is $1.6\%$.

The controllers of the approximants $f_4^*(t)$ are
$$
A_1=-0.99112\; , \qquad A_2=-0.77048 \; , \qquad A_3=-0.41456 \; ,
\qquad A_4=-0.10519\; ;
$$
$$
n_1=-0.17705\; , \qquad n_2=-0.05864\; , \qquad n_3=-0.05405\; ,
\qquad n_4=-0.06587 \; .
$$
The critical point and index are
$$
t_c=\frac{1}{|A_1|} = 1.0090\; , \qquad \al=|n_1|=0.177\; .
$$
The error of $t_c$ is $0.9\%$.

For $f_5^*(t)$, we obtain
$$
A_1=-0.9944\; , \qquad A_2=-0.8449 \; , \qquad A_3=-0.5782 \; ,
$$
$$
A_4=-0.2867\; , \qquad A_5=-0.0692\; ;
$$
$$
n_1=-0.1657\; , \qquad n_2=-0.0508 \; , \qquad n_3=-0.0438\; ,
$$
$$
n_4=-0.0452 \; , \qquad n_5=-0.0582 \; .
$$
The critical characteristics are
$$
t_c =\frac{1}{|A_1|} = 1.0056\; , \qquad \al=|n_1|=0.166 \; .
$$
Now, the error of $t_c$ is just $0.6\%$. Both the critical point and index
show good numerical convergence to exact results. The best Pad\'e approximant,
based on the same number of perturbative terms, is much less accurate giving
$t_c=1.117$ and $\al=1$; the error of $t_c$ being $12\%$.

\subsection{Critical indices from Wilson expansion}

The scalar quantum field theory with $\vp^4$-interaction is an ideal ground
for testing new methods in perturbation theory [2,48]. Much efforts has been
invested into studying critical phenomena by means of the $O(n)$-symmetric
$\vp^4$-theory. A very popular problem consists in calculating critical 
exponents by field theoretic renormalization group techniques in $d=4-\ep$ 
dimensions, resulting in the Wilson $\ep$-expansions [2,49--51]. To illustrate
a good accuracy of self-similar factor approximants in summing such 
$\ep$-expansions, we consider here the critical index $\nu$ for the 
correlation length $\xi(T)$, defined as 
\be
\label{23}
\nu\equiv -\lim_{T\ra T_c+0}\; (T-T_c)\frac{d}{dT}\; \ln\xi(T) \; ,
\ee
where $T$ is temperature. At the critical temperature $T_c$, the correlation
length diverges as $\xi\sim(T-T_c)^{-\nu}$. We shall use the expansion [49]
for the function
\be
\label{24}
f(\ep) \equiv \frac{1}{2\nu(\ep)} \qquad (\ep\equiv 4-d) \; .
\ee
From here, one can get the critical index in three dimensions
\be
\label{25}
\nu = \frac{1}{2f(1)} \qquad (d=3)
\ee
or in two dimensions
\be
\label{26}
\nu = \frac{1}{2f(2)} \qquad (d=2) \; .
\ee
Function (24), presented as expansion (2) in powers of $\ep$, in the
single-component case possesses the coefficients
$$
a_1=-0.1665\; , \quad a_2=-0.05865\; , \quad a_3=0.06225\; ,
\quad a_4=-0.1535\; , \quad a_5=0.4755\; .
$$
The straightforward usage of such expansions does not give reliable results, 
since the $\ep$-series are asymptotic and divergent. For instance, the 
critical index 
\be
\label{27}
\nu_k \equiv \frac{1}{2f_k(\ep)} \qquad (\ep\equiv 4-d) \; ,
\ee
defined through a $k$-order perturbative polynomial $f_k(\ep)$, as in Eq. (2),
for $d=3$ becomes
$$
\nu_1=0.6\; , \quad \nu_2=0.645\; , \quad \nu_3=0.597\; , \quad
\nu_4=0.731\; , \quad \nu_5=0.431 \qquad (d=3)\; .
$$
The sequence $\{\nu_k\}$, evidently, does not converge.

For the factor approximant $f_2^*(\ep)$, we obtain
$$
A_1=3.018517\; , \qquad A_2=-0.439949\; ;
$$
$$
n_1=6.874957\times 10^{-3} \; , \qquad n_2=0.425622\; .
$$
The corresponding index
\be
\label{28}
\nu_k^* \equiv \frac{1}{2f^*_k(\ep)} \qquad (\ep\equiv 4-d)
\ee
in $\nu_2^*=0.634$ for $d=3$ and $\nu_2^*=1.216$ for $d=2$.

We know that for the two-dimensional Ising model, when $\ep=2$, the 
index $\nu=1$. Imposing this restriction, we find for $d=3$ the index 
$\nu_2^*=0.629$. This is to be compared with the Borel-transform based 
calculations for the $\ep$-expansion, yielding $\nu=0.628\pm0.001$ [52] 
and $\nu=0.629\pm0.003$ [53] and also with the lattice numerical results 
$\nu=0.631\pm 0.002$ [2].

\subsection{Critical characteristics from high-temperature expansions}

High-temperature series expansions are often used in studying critical 
phenomena. Here, we shall consider the so-called $(2+1)$-dimensional 
Ising model, whose Hamiltonian reads
\be
\label{29}
H =\sum_i \left ( 1 -\sgm_i^z\right ) - g \sum_{<ij>} \sgm_i^x \sgm_j^x -
h \sum_i \sgm_i^x \; ,
\ee
in which $\sgm_i^\al$ are Pauli matrices; $g$, coupling parameter; $h$, 
magnetic field; summation is over a $2$-dimensional spatial lattice; 
$<ij>$ denotes nearest-neighbour pairs of sites. This model is in the same 
universality class as the three-dimensional Ising model. High-temperature 
expansions for this model are equivalent to the weak-coupling expansions in 
powers of $g$.

Consider, first, the mass gap
\be
\label{30}
F(g) \equiv E_1(g) - E_0(g) \; ,
\ee
being the energy difference between the first excited and ground state 
levels, at zero magnetic field. For the reduced quantity
\be
\label{31}
f(g) \equiv \frac{F(g)}{F(0)} \; , \qquad F(0)=2\; ,
\ee
the series expansions in powers of $g$ are of type (2).

The coefficients in series (2) for a square lattice are [54]
$$
a_1 = -2\; , \qquad a_2 = -1\; , \qquad a_3 =-\; \frac{3}{2}\; ,
\qquad a_4 = -\frac{9}{4}\; , \qquad a_5 =- \; \frac{11}{2}\; ,
$$
$$
a_6 = -10.253906\; , \qquad a_7 =-28.849609\; , \qquad a_8 =-57.418152\; ,
$$
$$
a_9=-175.053360\; , \qquad a_{10}=-365.267989\;  .
$$
When $g$ approaches the critical value $g_c$, the mass gap tends to zero, as
$F\sim (g_c-g)^\nu$, with the critical index
\be
\label{32}
\nu \equiv -\lim_{g\ra g_c-0}\; (g_c-g)\frac{d}{dg}\; \ln f(g) \; .
\ee
Note that the coefficients $a_i$ in series (2), for the present case, are of 
the same sign, which is known to be a difficult situation for summation. The 
absolute values of these coefficients are quickly increasing, which manifests 
the series divergence.

For the self-similar factor approximant $f_2^*(g)$, we find
$$
A_1=4.03553\; , \qquad A_2=-3.03553\; ; 
$$
$$
n_1=-2.4905\times 10^{-3} \; , \qquad n_2=0.65555\; .
$$
The critical coupling and critical index are
$$
g_c =\frac{1}{|A_2|} = 0.32943 \; , \qquad \nu=n_2=0.656 \; .
$$

For the factor approximant $f_3^*(g)$, we have
$$
A_1=-3.03055\; , \qquad A_2=2.34432+1.17108\; i \; ; \qquad A_3=A_2^*\; ;
$$
$$
n_1=0.65826 \; , \qquad n_2=(1.57037+5.33043\; i)\times 10^{-3} \; ,
\qquad n_3=n_2^* \; ,
$$
which gives the critical characteristics
$$
g_c =\frac{1}{|A_1|} =0.32997 \; , \qquad \nu=n_1=0.658 \; .
$$

Approximant $f_4^*(g)$ possesses
$$
A_1=-3.03764\; , \qquad A_2=2.80311\; , \qquad A_3=-0.85714+1.50878\; i \; ,
\qquad A_4=A_3^*\; ;
$$
$$
n_1=0.6514 \; , \quad n_2=-7.37667 \times 10^{-3} \; , \quad
n_3=(-7.68487+4.56213\; i)\times 10^{-3} \; ,  \quad n_4=n_3^* \; .
$$
This results in
$$
g_c = \frac{1}{|A_1|} =0.32920 \; , \qquad \nu=n_1=0.651 \; .
$$

Finally, for $f_5^*(g)$, we obtain
$$
A_1=-3.04301\; , \qquad A_2=2.85769 \; , \qquad A_3=-2.08221 \; ,
$$
$$
A_4=0.05019+1.48535\; i\; , \qquad A_5=A_4^* \; ;
$$
$$
n_1=0.64147\; , \qquad n_2=-6.64215\times 10^{-3} \; , \qquad n_3=0.02454 \; ,
$$
$$
n_4=(-1.27747+ 7.47775\; i)\times 10^{-3} \; , \qquad n_5=n_4^* \; .
$$
The corresponding critical characteristics are
$$
g_c= \frac{1}{|A_1|} =0.32862 \; , \qquad \nu=n_1=0.641 \; .
$$
The found results can be compared with other calculations, summarized 
in Ref. [54], where the critical coupling is in the interval $0.328\leq 
g_c\leq 0.329$, while the critical index lies in the range 
$0.629\leq\nu\leq 0.646$. This is in good agreement with our estimates.

Let us also consider the susceptibility
\be
\label{33}
\chi(g) \equiv -\; \frac{1}{N}\; 
\frac{\prt^2 E_0}{\prt h^2}{\Big |}_{h=0} \; ,
\ee
which behaves as $\chi(g)\sim(g_c-g)^{-\gm}$ in the vicinity of the critical 
coupling $g_c$, with the critical index
\be
\label{34}
\gm \equiv \lim_{g\ra g_c-0}\; (g_c -g) \; \frac{d}{dg}\; \ln \chi(g) \; .
\ee
Expanding the susceptibility (33) in powers of $g$, one gets the series of 
type (2). For the triangular lattice the coefficients are
$$
a_1 = 6\; , \quad a_2 = 32.95\; , \quad a_3 =166.5\; ,
\quad a_4 = 843.046875\; , \quad a_5 =4218.416666\; ,
$$
$$
a_6 = 20941.023004\; , \qquad a_7 =103361.512587\; , 
\qquad a_8 =507986.371687\; ,
$$
$$
a_9=2488222.50870\; , \qquad a_{10}=12155136.2137\;  .
$$
Again, all coefficients are of the same sign and fastly increasing in their 
values. Such a series is divergent for any finite $g$.

For the factor approximant $\chi_2^*(g)$, we have
$$
A_1=6.22771\; , \qquad A_2= -4.58786\; ; 
$$
$$
n_1=-0.03523 \; , \qquad n_2=-1.35562\; .
$$
The critical coupling and index are
$$
g_c =\frac{1}{|A_2|} =0.21797 \; , \gamma=|n_2|=1.356 \; .
$$

In the next order, we get $\chi_3^*(g)$, with the controllers
$$
A_1=-4.81702\; , \qquad A_2=1.6542+2.20379\; i \; ; \qquad A_3=A_2^*\; ;
$$
$$
n_1=-1.18552 \; , \qquad n_2=0.22051+0.09988\; i \; , \qquad n_3=n_2^* \; .
$$
From here,
$$
g_c =\frac{1}{|A_1|}=0.20760 \; , \qquad \gamma=|n_1|=1.185 \; .
$$

For $\chi_4^*(g)$, we find
$$
A_1=-4.76379\; , \quad A_2=-0.89137\; , \quad 
A_3=0.79806+1.56616\; i \; , \quad A_4=A_3^*\; ;
$$
$$
n_1=-1.25533 \; , \qquad n_2=2.32460  \; , \qquad 
n_3=1.06769-0.12380\; i \; , \qquad n_4=n_3^* \; ,
$$
which yields
$$
g_c = \frac{1}{|A_1|} =0.20992 \; , \qquad \gamma=|n_1|=1.255 \; .
$$

And the approximant $\chi_5^*(g)$ contains
$$
A_1=-4.76723\; , \qquad A_2=-2.66216 \; , \qquad A_3=-1.30481 \; ,
$$
$$
A_4=0.84329+1.55612\; i\; , \qquad A_5=A_4^* \; ;
$$
$$
n_1=-1.24715\; , \qquad n_2=-0.10924 \; , \qquad n_3=1.54693 \; ,
$$
$$
n_4=1.02499-0.01717\; i \; , \qquad n_5=n_4^* \; .
$$
The corresponding critical characteristics are
\be
\label{35}
g_c = \frac{1}{|A_1|} =0.20977 \; , \qquad \gamma=|n_1|=1.247 \; .
\ee
These results are in good agreement with numerical calculations, summarized in
Ref. [54], according to which the critical coupling for the triangular lattice
is in the interval $0.20972\leq g_c\leq 0.20980$ and the critical index is in
the range $1.236\leq\gamma\leq 1.250$.

\section{General discussion and possible variants}

The self-similar factor approximants, as is illustrated above by a variety 
of examples, provide a powerful tool for defining effective sums of divergent 
power series. The latter in the majority of applications are asymptotic. 
However, the presented method is not limited to only asymptotic series. The 
sole thing that is required is that the perturbative form (2) be a finite 
{\it power} series. If all coefficients $a_n$ in Eq. (2) do not depend on 
the approximation order $k=1,2,\ldots$, then the difference
$$
\Delta f_k(x) \equiv f_k(x) - f_{k-1}(x)
$$
composes as asymptotic sequence $\{\Delta f_k(x)\}$, where $\Delta f_k(x)=
a_k\;x^k$. But we also may consider the case [5], when an iterative procedure 
results in the approximations
$$
f_k(x) = 1 + \sum_{n=1}^k a_{nk}\; x^n \; ,
$$
with different coefficients $a_{nk}\neq a_{n\; k-1}$. In such a situation,
the difference $\Delta f_k(x)$ does not constitute an asymptotic sequence
$\{\Delta f_k(x)\}$. Nevertheless, the method of self-similar factor 
approximants is applicable to this case as well.

The mathematical structure of formula (5) resembles, in some sense, such 
geometric structures as fractals [55,56]. With regard to functions, one 
usually says that a function possesses fractal properties if it manifests 
a kind of scaling. In particular, it can be an asymptotic scaling [56,57], 
arising in a limit of the function variable. In that sense, formula (5) 
displays the property of {\it asymptotic scaling}
\be
\label{36}
\lim_{x\ra\infty} \; \frac{f_k^*(\lambda x)}{f_k^*(x)} =
 \lambda^{\nu_k}\; , \qquad \nu_k = {\sum_{i=1}^k \; n_i} \; ,
\ee
where $\lambda$ is a scaling parameter. We also may notice another scaling 
property of the factor form (5). To this end, introducing functions
\be
\label{37}
\vp_i(x) \equiv (1 + A_i\; x)^{n_i} \; ,
\ee
we can present Eq. (5) as
\be
\label{38}
f_k^*(x) = \prod_{i=1}^k \vp_i(x) \; .
\ee
Then, the property of {\it functional scaling}
\be
\label{39}
f_k^*(x) \ra \left ( \prod_{i=1}^k \lambda_i\right ) \; f_k^*(x) \; ,
\qquad \vp_i(x) \ra \lambda_i\; \vp_i(x) \; ,
\ee
is valid.

It is worth emphasizing that the factor approximant (5) is principally 
different from the continued-function representation mentioned by Bender 
and Orszag [58]. According to their scheme, approximants $\bt_k(x)$ for 
a function $f(x)$ could be constructed in the following way. Let us take 
an arbitrary function $\bt(c,x)$, where $c$ is a parameter. Then a 
$k$-order approximation to the sought function $f(x)$ is given by
$$
\bt_k(x) \equiv \bt \left (
c_1,x\bt\left (c_2,\ldots,x\bt(c_k,x) \right ) \ldots \right ) \; . 
$$
However, such a manner of acting has several essential drawbacks. First 
of all, this scheme is too much ambiguous, since there exists an infinity 
of functions that could be accepted as $\bt(c,x)$. Contrary to this, the 
form of the factor approximants (5) is not arbitrary postulated but is 
derived from the given perturbative series by means of the self-similar 
approximation theory. Second, Eq. (5) does not have the form of a continued 
function $\bt_k(x)$. Really, even if we suppose that $\bt(c,x)=(1+cx)^n$, 
then 
$$
\bt_k(x) = \left ( 1 + c_1 x\left ( 1 + c_2x (1 + \ldots + c_k\; x)^n
\right )^n \ldots \right )^n \; ,
$$
which has nothing to do with formula (5). The latter is not a function of
the same function but a product (38) of different functions (37). This 
product is not only different from the above $\bt_k(x)$ but also is more 
general. Last but not least, the sequence $\{\bt_k(x)\}$ of the 
continued-function representation is, generally, speaking, {\it divergent}. 
This can be easily demonstrated for the above mentioned case of $\bt(c,x)$ 
resulting in the power-law representation $\bt_k(x)$. The latter, at large 
$x$, are of the type
$$
\bt_k(x) \simeq \left ( \prod_{i=1}^k c_i\right )^n \; x^{nk} \qquad
(x\ra\infty) \; ,
$$
which shows that the subsequent approximations $\bt_k(x)$ differ significantly
from each other. Thus, $\bt_1(x)\sim x^n$, $\bt_2(x)\sim x^{2n}$, and so on,
each $\bt_k(x)\sim x^{nk}$ having principally different dependence on $x$.
It is evident that such a sequence cannot converge.

The self-similar factor approximant (5) of order $k$ requires the knowledge
of $2k$ nontrivial (except the first term equal to one) perturbative terms of
series (2). Thus, for constructing $f_k^*(x)$, we need an even number of terms.
A natural question would be: how could we use an odd number of such terms?
There are two straightforward ways of doing this, which can be called additive
and multiplicative.

The {\it additive} way is as follows. The perturbative form (2) can always be
rewritten as
\be
\label{40}
f_k(x) = 1 + a_1 x\prod_{i=1}^{k-1} (1 + c_i\; x) \; .
\ee
By applying the self-similar renormalization (4) to the factors in the product
of Eq. (40), we come to the approximant
\be
\label{41}
f_{k+}^*(x) \equiv 1 + a_1x\prod_{i=1}^k (1 +B_i\; x)^{m_i} \; ,
\ee
containing the additive unitary term. The approximant (41) requires the
availability of an odd number of terms in series (2). If the large-$x$ 
behaviour (6) is known, it can be employed for approximant (41) as the 
crossover condition \be
\label{42}
a_1\; \prod_{i=1}^k B_i^{m_i}  = f_\infty \; , \qquad
\sum_{i=1}^k {m_i+1} =\bt \; .
\ee
We have analysed the usage of approximants (41) to the examples considered 
above. Sometimes these approximants $f_{k+}^*(x)$ are not bad, being of 
comparable accuracy with $f_k^*(x)$ or even better. This, e.g., is the case 
for defining the critical index (23) from the Wilson $\ep$-expansion, for 
which we find $\nu_{2+}^*=0.628$ for $d=3$ and $\nu_{2+}^*=0.914$ for $d=2$. 
However, for the majority of cases, the accuracy of $f_{k+}^*(x)$ is 
essentially worse than that of $f_k^*(x)$. On occasion, this accuracy can be 
improved by considering the inverse function $f^{-1}(x)$, instead of $f(x)$. 
This, for instance, gives good results for the structure factors of Section 4. 
The crossover condition (42), if available, also may certainly cure the 
behaviour of $f_{k+}^*(x)$. For example, this yields good approximations 
for the ground-state energy of the anharmonic oscillator of Section 5.2. 
But, in general, the approximants $f_{k+}^*(x)$ are inferior to $f_k^*(x)$.

As an alternative, we could try the {\it multiplicative} way of extracting
information from odd numbers of perturbative terms. For this purpose, we may
present Eq. (2) as
\be
\label{43}
f_k(x) = (1+Ax) \prod_{i=1}^{k-1} ( 1 + b_i\; x) \; .
\ee
Accomplishing the self-similar renormalization of the right-hand-side factors
in Eq. (43), except the first factor $(1+Ax)$, we define the self-similar
approximant
\be
\label{44}
f_{k+1/2}^*(x) \equiv (1 + Ax) \; \prod_{i=1}^k (1 +A_i\; x)^{n_i}
\ee
In the starting trivial case,
\be
\label{45}
f_{1/2}^*(x) \equiv f_0^*(x) = 1+ a_1 x\; .
\ee
The following nontrivial approximants are
$$
f_{3/2}^*(x) = (1 +Ax)(1+A_1x)^{n_1} \; , 
$$
\be
\label{46}
f_{5/2}^*(x) = (1+Ax)(1+A_1x)^{n_1}(1+A_2x)^{n_2} \; ,
\ee
and so on. Each $f_{k+1/2}^*(x)$ requires for its definition $2k+1$ terms of
series (2).

The main difference in employing approximants $f_{k+1/2}^*(x)$, as compared 
to $f_k^*(x)$, is in the following. When determining the related control 
parameters by means of the order-through-accuracy matching, for the former 
approximants we may get nonunique solutions, while for the latter the 
solutions are unique. Different solutions for the control parameters yield 
several possible variants of $f_{k+1/2}^*(x)$. As we have checked, all of 
these several variants, generally, have reasonable accuracy, being close 
to each other, However, dealing with several approximants of the same order 
complicates the calculational procedure. In this way, the self-similar 
factor approximants $f_k^*(x)$ look to be more preferable, compared to both 
$f_{k+}^*(x)$ and $f_{k+1/2}^*(x)$.

In all cases, we have considered, the accuracy of $f_k^*(x)$ improves with 
the increasing order $k$. This implies the existence of the so-called {\it 
numerical convergence} for the sequence $\{ f_k^*(x)\}$. It is, of course, 
impossible to guarantee that the latter sequence always converges to the 
exact function $f(x)$. This kind of uncertainty, is actually, common for 
practically any sequence of numerical approximants. To render the procedure 
uniquely defined requires imposing on the calculational algorithm additional 
conditions making the latter single-valued [59].

More information of the self-similar factor approximants $f_k^*(x)$ can 
be extracted by invoking the probabilistic analysis [60]. To this end, we 
introduce the multiplier 
\be
\label{47}
M_k(x) \equiv \frac{\dlt f_k^*(x)}{\dlt f_0^*(x)} \; .
\ee
Taking account Eq. (8), we get
\be
\label{48}
M_k(x) = \frac{1}{a_1}\; \frac{d}{dx}\; f_k^*(x) \; .
\ee
The probability of $f_k^*(x)$ is defined as
\be
\label{49}
p_k(x) = \frac{|M_k^{-1}(x)|}{\sum_{i=1}^N\; |M_i^{-1}(x)|} \; ,
\ee
where $N$ is the total number of available approximants $f_k^*(x)$, with
$k=1,2,\ldots,N$. Having defined these quantities, we now may find the most 
probable approximants, corresponding to the maximal probability (49). 
Also, we can calculate a {\it weighted approximant}
\be
\label{50}
\overline f_k(x) \equiv \sum_{j=1}^k p_j(x)\; f_j^*(x) \; .
\ee
This averaging is especially useful when the subsequent approximants 
$f_j^*(x)$ and $f_{j+1}^*(x)$ display large variations. The latter become 
smoothed by the averaging procedure (50). For example, for the critical 
index of the correlation length, discussed in section 6.2, defined by 
this averaging procedure, we find $\overline\nu=0.630$.

Additional possibilities arise, when, for the same problem, there exist 
expansions for several different functions. Say, for simplicity, there
are two functions $f(x)$ and $g(x)$, for which the expansions $f_k(x)$ and
$g_k(x)$ are available at $x\ra 0$. Then, in addition to the factor 
approximants $f_k^*(x)$ and $g_k^*(x)$, we may, considering some combinations
$F_k(f_k(x),g_k(x))$ of the series $f_k(x)$ and $g_k(x)$, construct the
approximants $F_k^*(x)$. For instance, we can consider the sums 
$f_k(x)+g_k(x)$, products $f_k(x)g_k(x)$, or the ratios $f_k(x)/g_k(x)$,
or other combinations of the given series. A more detailed consideration
of such possibilities will be done in future.

Finally, we  would like to emphasize that the accuracy of the self-similar 
factor approximants (5), for the examples considered in the present paper, 
is always essentially better than that of the best Pad\'e approximants, 
constructed with the same number of perturbative terms of series (2). This 
fact is easy to understand, since the factor approximants (5) have a much 
more general form as compared to Pad\'e approximants, including the latter 
just as a narrow particular class. As is explained in Section 2, the 
self-similar factor approximants can represent rational, irrational,
and transcendental functions.

\vskip 5mm

{\bf Acknowledgement}

\vskip 2mm

One of us (V.I.Y.) is grateful to E.P. Yukalova for useful discussions 
and advice.

\newpage

\begin{center}
{\large{\bf Figure Captions}}
\end{center}

\vskip 1cm

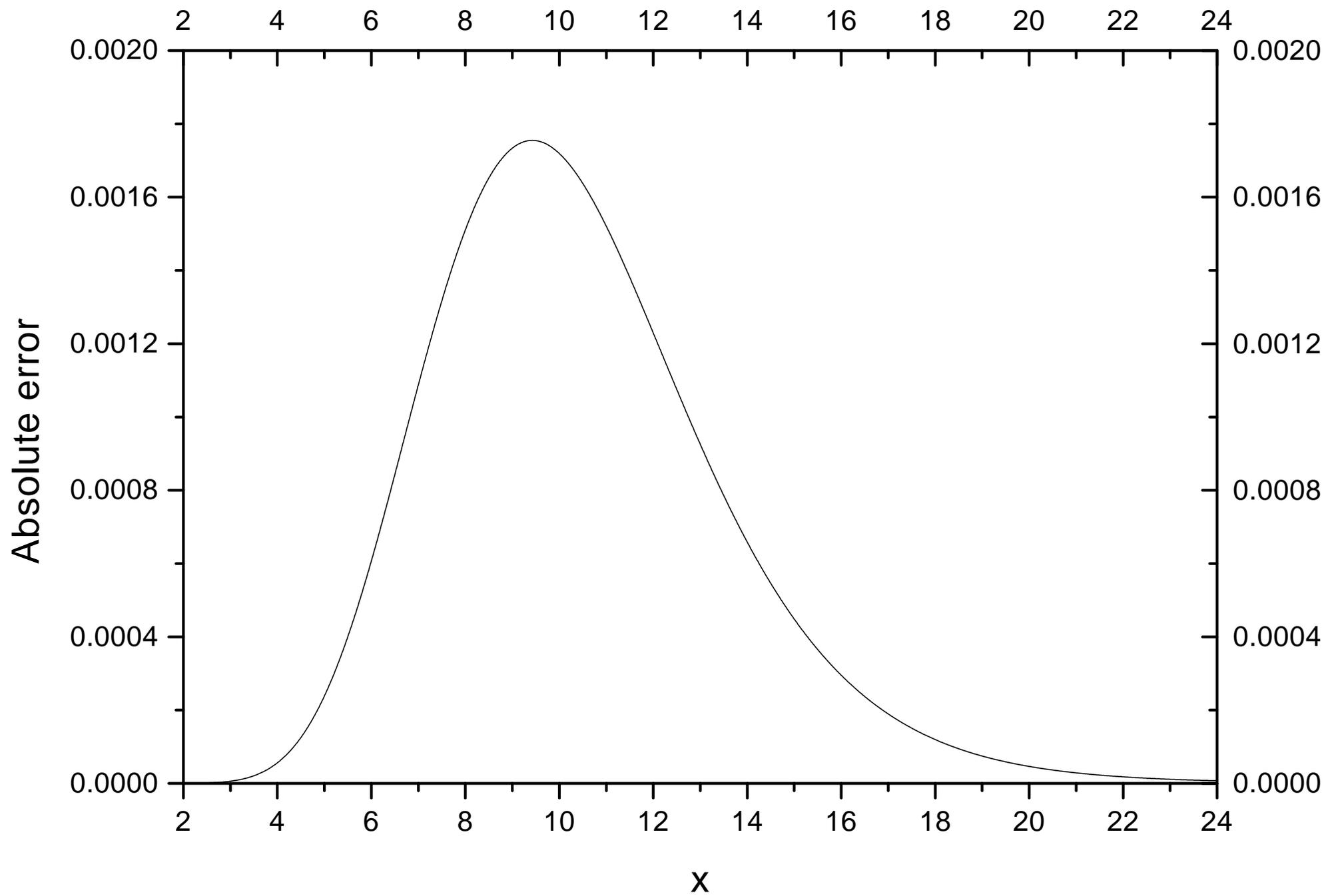
{\bf Fig. 1}. Difference of the self-similar factor approximant $f_4^*(x)$ 
and function (13).

\vskip 1cm

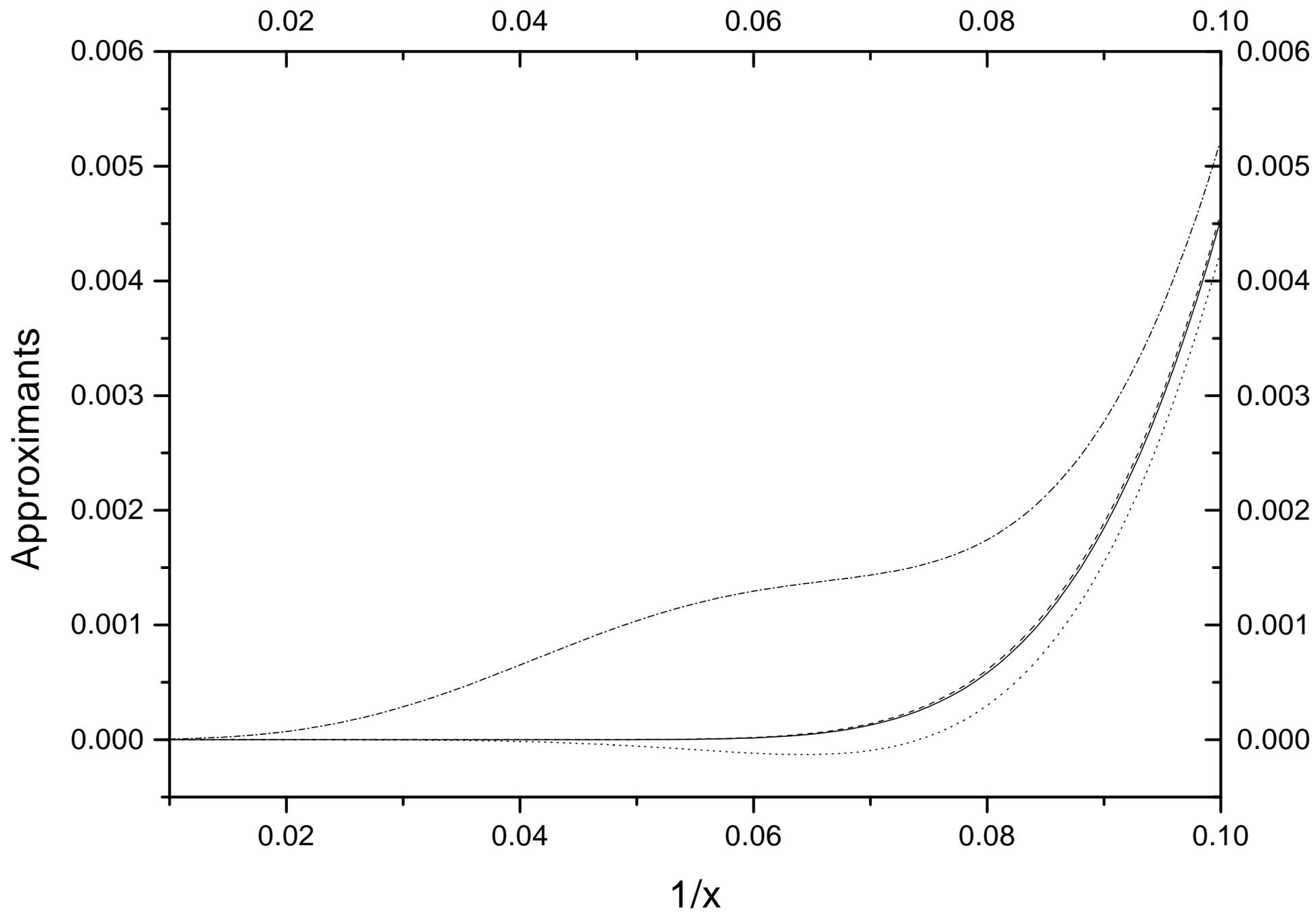
{\bf Fig. 2}. Specific heat (14) (solid line), the approximant $C_3^*(x)$ 
(dashed line), and the best Pad\'e approximants $P_{[4/2]}(x)$ (dashed-dotted 
line) and $P_{[5/1]}(x)$ (dotted line) versus $1/x$.

\vskip 1cm

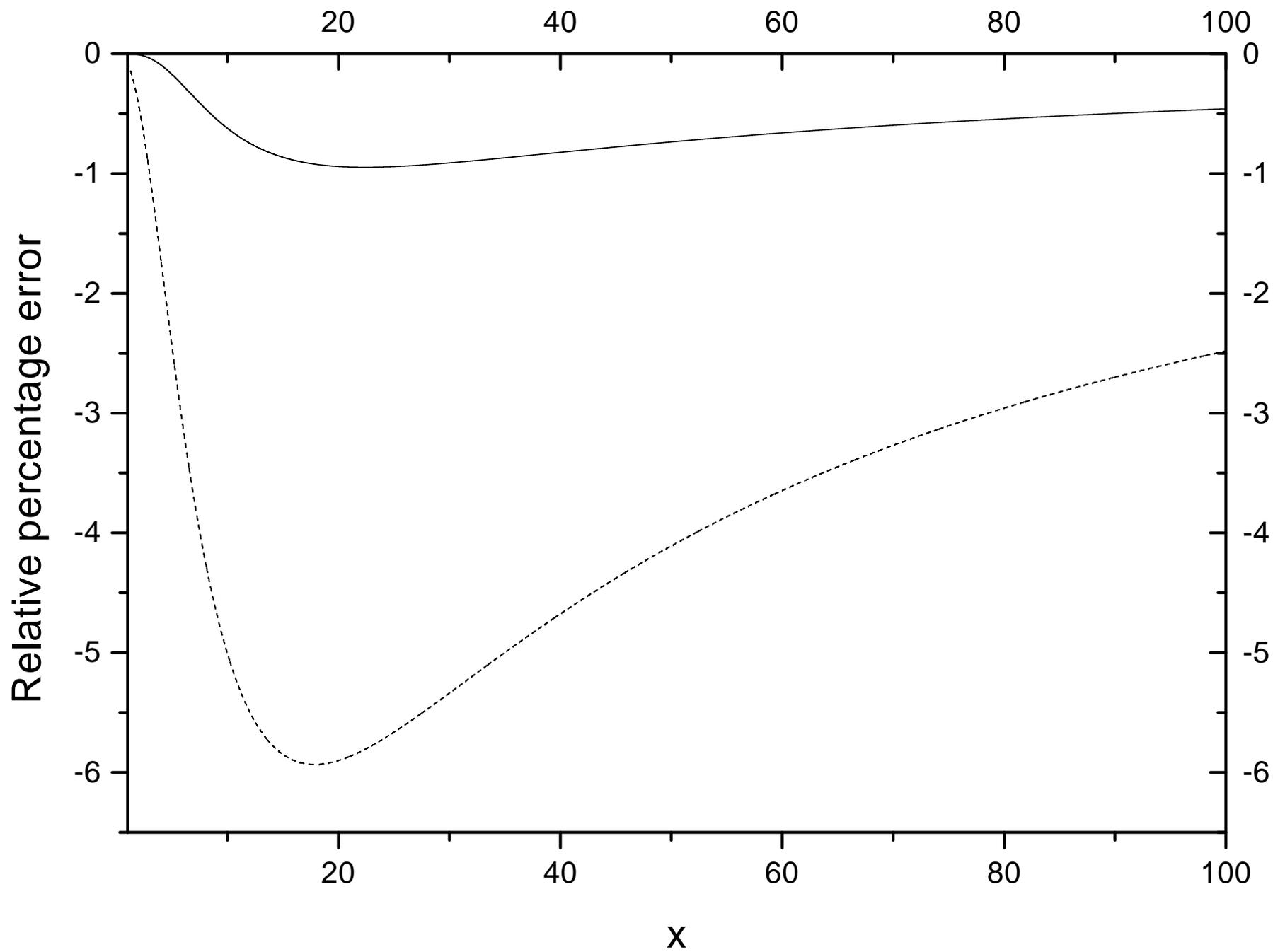
{\bf Fig. 3}. Percentage errors of the approximant $f_3^*(x)$ (solid line) 
and the two-point Pad\'e approximant $P_{[3/2]}(x)$ (dashed line), as
compared to the exact function (15).

\vskip 1cm

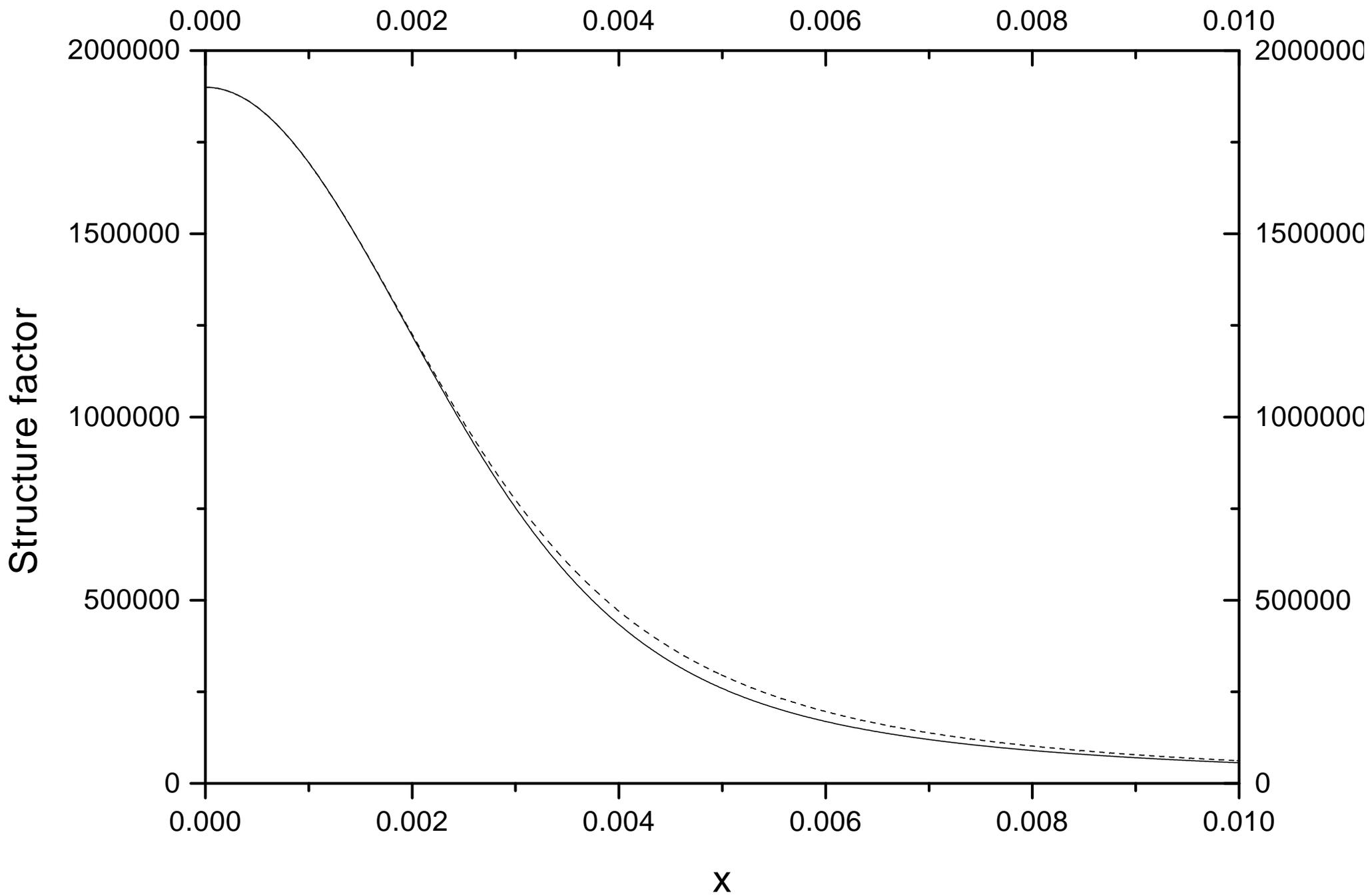
{\bf Fig. 4}. Structure factor of branched polymer $S(x)$ (solid line) 
and its factor approximant $S_2^*(x)$ (dashed line).

\vskip 1cm

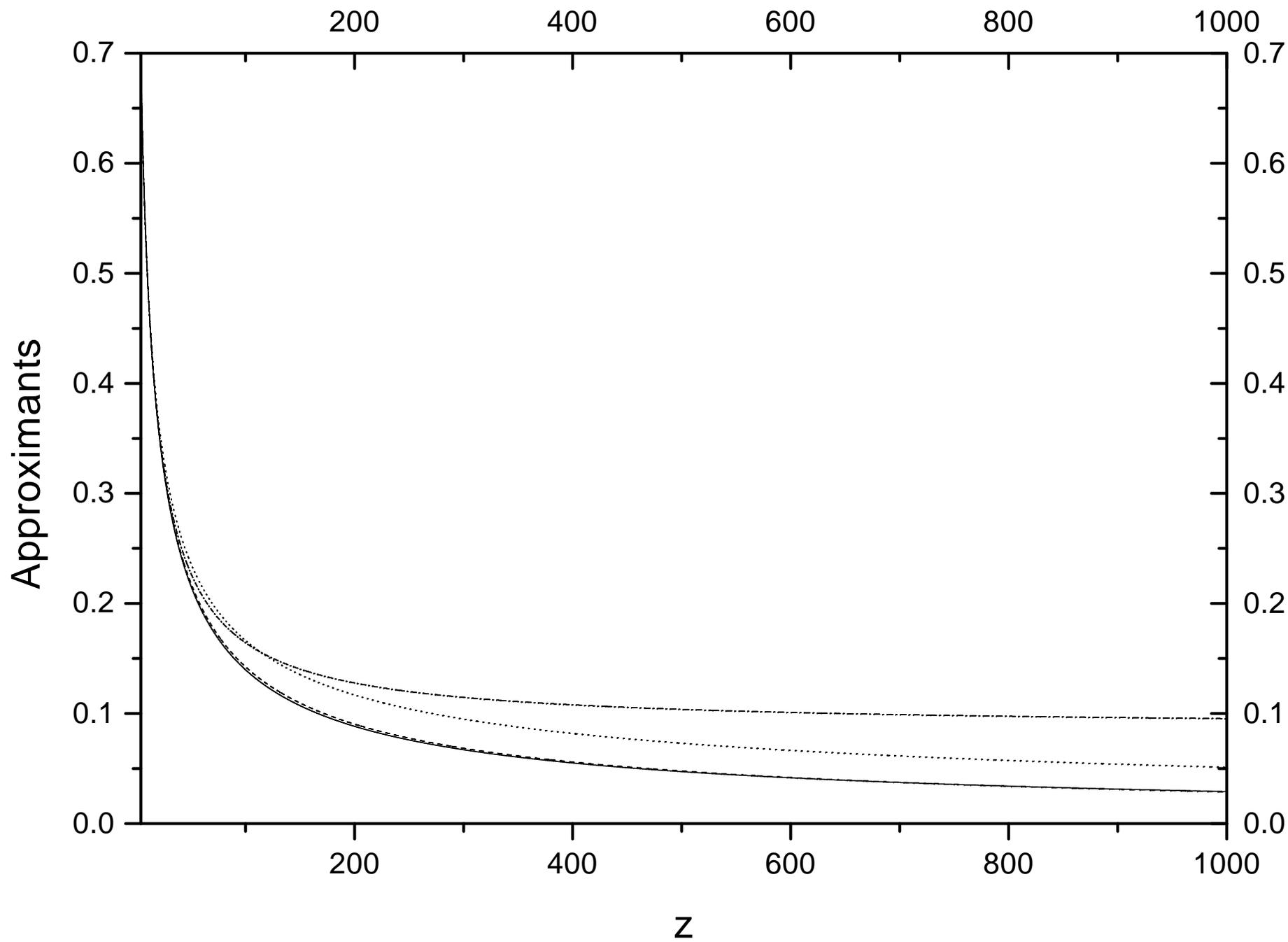
{\bf Fig. 5}. Luminiscent intensity $f(z)$ (solid line), approximant 
$f_3^*(z)$ (dashed line) and the best Pad\'e approximant $P_{[3/3]}(x)$
(dashed-dotted line).

\end{document}